\documentclass[pdflatex,sn-basic]{sn-jnl}

\usepackage{lineno,hyperref}
\usepackage{anyfontsize}
\usepackage{adjustbox}
\usepackage{longtable}
\usepackage{xcolor}
\usepackage{multirow}
\usepackage{multicol}
\usepackage{tabu}
\usepackage{tocloft}
\usepackage{blindtext}
\usepackage{mathtools}
\usepackage{blkarray}

\renewenvironment{table}[1][]%
{\tableorg[#1]%
\tablebodyfont%
\renewcommand\footnotetext[2][]{{\removelastskip\vskip3pt%
\let\tablebodyfont\tablefootnotefont%
\hskip0pt\if!##1!\else{\smash{$^{##1}$}}\fi##2\par}}%
}{\endtableorg}

\usepackage{float}
\usepackage{rotating}
\usepackage{lscape}
\usepackage{array}
\usepackage{booktabs}
\usepackage{soul}
\usepackage{textcomp}
\usepackage{amsmath}
\usepackage{amssymb}
\usepackage{setspace}
\usepackage{bm}
\usepackage{makecell}

\DeclareUnicodeCharacter{2212}{ }

\usepackage{mathrsfs}
\jyear{}

\theoremstyle{thmstyleone}

\theoremstyle{thmstyletwo}%

\theoremstyle{thmstylethree}%

\begin{document}

\title[Partial membership models for soft clustering of multivariate football data]{Partial membership models for soft clustering of multivariate football player performance data}

\author*[1]{\fnm{Emiliano} \sur{Seri}}\email{emiliano.seri@uniroma2.it}

\author[2]{\fnm{Roberto} \sur{Rocci}}\email{roberto.rocci@uniroma1.it}

\author[3]{\fnm{Thomas Brendan} \sur{Murphy}}\email{brendan.murphy@ucd.ie}

\affil[1]{\orgdiv{Department of Enterprise Engineering “Mario Lucertini”
}, \orgname{University of Rome Tor Vergata}, \orgaddress{Via del Politecnico 1, 00133, Rome, Lazio, Italy}}

\affil[2]{\orgdiv{Department of Statistics}, \orgname{Sapienza University of Rome}, \orgaddress{Piazzale Aldo Moro 5,
00185, Rome, Lazio, Italy}}

\affil[3]{\orgdiv{School of Mathematics and Statistics}, \orgname{University College Dublin}, \orgaddress{Dublin, D04 V1W8, Ireland}}

\abstract{
The standard mixture modeling framework has been widely used to study heterogeneous populations, by modeling them as being composed of a finite number of homogeneous sub-populations. However, the standard mixture model assumes that each data point belongs to one and only one mixture component, or cluster, but when data points have fractional membership in multiple clusters this assumption is unrealistic. It is in fact conceptually very different to represent an observation as partly belonging to multiple groups instead of belonging to one group with uncertainty. For this purpose, various soft clustering approaches, or individual-level mixture models, have been developed.  In this context, \cite{heller2008statistical} formulated the Bayesian partial membership model (PM) as an alternative structure for individual-level mixtures, which also captures partial membership in the form of attribute-specific mixtures. 

Our work proposes using the PM for soft clustering of count data arising in football performance analysis and compares the results with those achieved with the mixed membership model and finite mixture model. Learning and inference are carried out using Markov chain Monte Carlo methods. 

The method is applied on Serie A football player data from the 2022/2023 football season, to estimate the positions on the field where the players tend to play, in addition to their primary position, based on their playing style.
The application of partial membership model to football data could have practical implications for coaches, talent scouts, team managers and analysts. These stakeholders can utilize the findings to make informed decisions related to team strategy, talent acquisition, and statistical research, ultimately enhancing performance and understanding in the field of football.}


\maketitle

\section{Introduction}
\label{sec:intro}

Model-based clustering has been widely used among researchers to study heterogeneous populations, by modeling them as being composed of a finite number of homogeneous sub-populations \citep[eg.][]{fraley2002model, peel2000finite, Bouveyron2019Jun, gormley2023model}.
In this framework, observations within a dataset are modeled as originating from one of multiple probability distributions. 
The objective is to achieve a clustering solution where observations are partitioned into distinct groups. Notably, observations that show a significant posterior probability of belonging to more than one cluster are considered to have uncertain group membership. This uncertainty can sometimes suggest a model that does not adequately fit the data.
However, the standard mixture model assumes that each data point belongs to one and only one mixture component or cluster, but when data points have fractional membership in multiple clusters this assumption is unrealistic. 
In contrast,  mixed membership and partial membership models accommodate partial membership to multiple clusters. 
In a standard finite mixture model, if an observation appears to have a 0.5 probability of belonging to two clusters, this is interpreted as uncertainty about which single cluster is correct. By contrast, in a partial membership model, such a 50/50 split indicates that the observation is genuinely comprised of both clusters to an equal degree, rather than being purely one or the other. This conceptual difference is central to the idea of partial membership.
For example, let's consider a football player with the role of playmaker. 
His duties on the pitch change depending on the phase of play. In an offensive phase, they are those associated with the attacking role, while in a defensive phase they are those associated with the midfielder role.
That football player should be represented as partly belonging to two different classes or sets.
Being certain that a player's role is partly midfielder and partly striker is very different from being uncertain about a player's role on the pitch.

The foundational concept of mixed membership modeling traces its origins to the 1970s with the development of the Grade of Membership (GoM) model by mathematician Max Woodbury, which was designed for “fuzzy” classifications in medical diagnostic scenarios \citep{woodbury1978mathematical}.
It was not until the early 2000s, following the widespread adoption of Bayesian methods and an enhanced understanding of the duality between the discrete and continuous nature of latent structures in the GoM model, that a new Bayesian approach to the GoM model was developed. Independently, and within a short span of time, three mixed membership models emerged to address challenges in distinctly different domains: \cite{blei2003latent} -- Latent Dirichlet Allocation (LDA);  \cite{erosheva2003bayesian} -- Grade of Membership model (GoM); \cite{pritchard2000association} -- Admixture model.
Mixed membership models unify the LDA, GoM, and admixture models in a common framework and provides ways to construct other individual-level mixture models by varying assumptions on the population, sampling unit and latent variable levels, and the sampling scheme.

Partial membership models  \citep{heller2008statistical} are defined using a similar framework, but they overcome some of the drawbacks of mixed membership models, such as the assumption of conditional independence between attributes and the challenges associated with interpreting cluster memberships in complex datasets. By relaxing these assumptions, partial membership models provide a more flexible and interpretable framework, particularly in applications with nuanced relationships between attributes, such as football player performance analysis.
While \cite{heller2008statistical} focus on continuous data and a more general setup, in the present paper, we build on the partial membership framework by adapting it specifically for count data and incorporating independent Poisson likelihoods (with Gamma priors) for each variable within a cluster. This approach is well-suited to discrete, nonnegative observations. We also address potential identifiability concerns by leveraging “archetypal units,” i.e., observations whose membership vector $\bm{\tau}_i$ is very close to 1 for one cluster. By identifying real examples that typify each cluster’s core characteristics, this step facilitates both interpretability and stable estimation of partial memberships.

Even though partial membership models exist for some time, they are still under utilized in literature. 
\cite{gruhl2013tale} apply partial membership model to NBA player data from the 2010–11 season and compare the results with those achieved with the mixed membership model.
\cite{Chao2017image} used partial membership for semantic image segmentation, while \cite{hou2022chimeral} starts from partial membership models and epistatic clustering \citep{zhang2013epistatic} to develop a “hybrid method” between the two, to cluster hybrid species of flowers, which tend to exhibit a mixture of parent characteristics. An alternative approach to modeling multiple cluster structures is presented by \cite{GALIMBERTI2007520}, who propose a method that partitions variables into disjoint subsets, applying separate mixture models to each subset. This allows an observation to belong to different clusters based on varying subsets of variables, effectively capturing multiple clustering patterns within the data. In contrast, our Partial Membership model assigns each observation a membership vector $\tau$ over the entire set of variables, providing an integrated representation of partial memberships across all clusters simultaneously. This unified approach facilitates straightforward interpretation in contexts where a holistic understanding of cluster memberships is desired. \cite{GALIMBERTI2007520} method is instead useful when distinct clustering structures exist in different subsets of variables. Additionally, \cite{hou2022factor} propose a factor and hybrid components approach for Gaussian model-based clustering, which shares a conceptual resemblance to partial membership in that some parameters can be “hybridized” across components.

To demonstrate the practical utility of the extended PM model, we apply the PM model to Serie A football player performance data, from the 2022/2023 football season. The aim is to estimate the various roles players are inclined to occupy on the field, beyond their primary positions, by analyzing their playing styles. The goal of the application within sports analysis is to assist coaches, talent scouts, team managers, and analysts in making more informed decisions. These decisions pertain to team strategy, talent acquisition, and advancing statistical research in sports.
Another application presented in the Appendix~\ref{sec: applBike} is on the daily usage of bike sharing bikes in Washington, from June 15th to July 15th, 2022.
For both the bike-sharing and football scenarios, the variables of interest (number of rides originating at each station, number of shots, passes, tackles, etc.) are nonnegative integer counts with no intrinsic upper bound. The Poisson distribution is the most natural baseline model for count data under these conditions and is widely used across sports analytics and transportation research. While we acknowledge that real-world data may exhibit overdispersion or zero inflation in some contexts, the Poisson assumption provides a straightforward, interpretable starting point. Moreover, partial membership allows each observation to mix different Poisson “profiles,” thereby accommodating some variation that might otherwise lead to modest overdispersion. In cases where greater flexibility is required, extensions, such as negative binomial or zero-inflated formulations, could be explored in future work.

Mixed membership model for count data, outlined in \cite{white2016exponential}, and mixture model, are also applied and the results are compared. The comparison suggests that in this application, the partial membership model gives more realistic and interpretable results.

The article is organised as follows.
Section~\ref{sec: partial} outlines the modeling framework. From mixture model, to our Bayesian partial membership model specification with Poisson component distributions, and mixed membership model. It also compares the data generative process of the three models, and gives an overview on technical aspects like label switching and model selection, justifying the choice of the information criterion through literature and a simulation. In Section~\ref{sec: applBall}, partial membership model is applied to Serie A football players performance data and the results are compared with those achieved with the mixed membership mixture model. For the sake of completeness, a Poisson mixture model is also applied to this study, and the results are compared with those of the other two models, even if a crisp clustering approach does not align with the specific objectives and purposes of the proposed application.
Conclusions and future developments follow in Section~\ref{sec: conclusions}.
To further demonstrate the versatility and effectiveness of the proposed model, we include an additional application of the model to Washington DC bike sharing data in the Appendix~\ref{sec: applBike}.

\section{Modeling Framework}
\label{sec: partial}

\subsection{Mixture Model}

Consider a data set $\bm{X}=\{\bm{x}_{ij}:i=1, 2,\dots, N, j=1, 2,\dots, J\}$ with $N$ observations and $J$ features per observation.
In a finite mixture model the data are modeled as a mixture of $K$ component densities $P_k(\cdot\mid\bm{\theta}_k)$ with unknown mixing proportions ${\pi}_{1}, \dots, {\pi}_{K}$.
The density of $\bm{x}_{i}$ is given as:

\begin{equation*}
    P(\bm{x}_{i}\mid\bm{\Theta}, \bm{\pi})= \sum_{k=1}^{K}{\pi}_{k}P_{k}(\bm{x}_{i}\mid\bm{\theta}_{k}),
\end{equation*}
where the mixing proportion $\bm{\pi}=\{{\pi}_{k} : k=1, 2, \dots, K\}$ satisfy $\sum_{k=1}^{K}\pi_{k}=1$ with $\pi_{k}\geq 0$.
The component-specific parameters are collected in $\bm{\Theta}=\{\bm{\theta}_{k}: k=1,2,\ldots, K\}$.
In this framework, each data point $\bm{x}_i$ is assumed to come from one (and only
one) of the $K$ mixture components. Let
$\bm{\tau}_{i}$ be the latent component membership indicator variable for observation $i$, so $\tau_{ik}\in\{0,1\}$ and $\sum_{k}\tau_{ik}=1$. In the mixture model, we assume that $\bm{\tau}_{i}\sim \mbox{multinomial}(1, \bm{\pi})$, and so $\bm{\tau}_{i}=(\tau_{i1}, \dots, \tau_{iK})$ is the binary membership vector, where $\tau_{ik}=1$ indicates that the data point $\bm{x}_{i}$ was generated by mixture component $k$.

\subsection{Bayesian Partial Membership Model}
In a partial membership model, we relax the constraint $\tau_{ik}\in\{0,1\}$ to take any continuous value in the range $[0,1]$.  
$\tau_{i}=(\tau_{i1}, \tau_{i2},\dots,\tau_{iK})$ is the vector of partial membership weights for the $i$-th observation. These weights lie in the unit simplex, such that $\tau_{ik}\geq0$ and $\sum_k \tau_{ik}=1$.
In the partial membership model, the density of an observation becomes, 
\begin{equation*}
    P(\bm{x}_i\mid\bm{\Theta}) \propto \int_{\bm{\tau}_i}P(\bm{\tau}_{i})\prod_{k=1}^{K}P_{k}(\bm{x}_i\mid\bm{\theta}_k)^{\tau_{ik}}d\bm{\tau}_{i}.
\end{equation*}
We integrate over all values of $\bm{\tau}_i$ instead of summing.
This formulation allows for continuous membership across multiple clusters, contrasting with finite mixtures where membership is binary.
The product over $k$ reflects that the contribution to the likelihood of $\bm{x}_i$ from each component is compounded multiplicatively, representing how each component contributes to explaining $\bm{x}_{i}$ given its partial membership weights $\tau_{ik}$.
The exponent $\tau_{ik}$ on $P_k(\bm{x}_i\mid\bm{\theta}_k)$ represents the degree of membership of $\bm{x}_{i}$ in component $k$. This is not directly present in the finite mixture model, where the component membership is implicit in the mixing weights $\pi_k$ and assumed to be binary (fully in one component). In the partial membership context, $\tau_{ik}$ as an exponent softens this assumption, allowing $\bm{x}_{i}$ to “partially belong” to multiple components simultaneously, each to varying degrees expressed by $\tau_{ik}$.

The conditional density $\tau \mid x_i$ is a continuous distribution over the simplex, capturing the relative contributions of each component to explaining the observation $x_i$. For example, in a two-component mixture ($K=2$) with components $N(0,1)$ and $N(4,1)$, observing $x=2$ leads to a continuous posterior distribution over $\tau_{i}=(\tau_{i1}, \tau_{i2})$. A mode of $(\tau_{i1}, \tau_{i2})=(0.5,0.5)$ would indicate equal partial membership in both components. This is conceptually distinct from finite mixtures, where $\tau_{i}=(0,1)$ or $(1,0)$ reflects binary cluster membership with uncertainty.
The PM model thus enables nuanced interpretations of partial membership. For instance in the proposed application, a football player with $\tau=(0.5,0.5)$ can be understood as splitting their playing style equally between two roles, such as striker and midfielder. This interpretation reflects the simultaneous contribution of both clusters, rather than uncertainty about membership.

\begin{figure}[tbph]
    \centering
    \includegraphics[width=.8\linewidth]{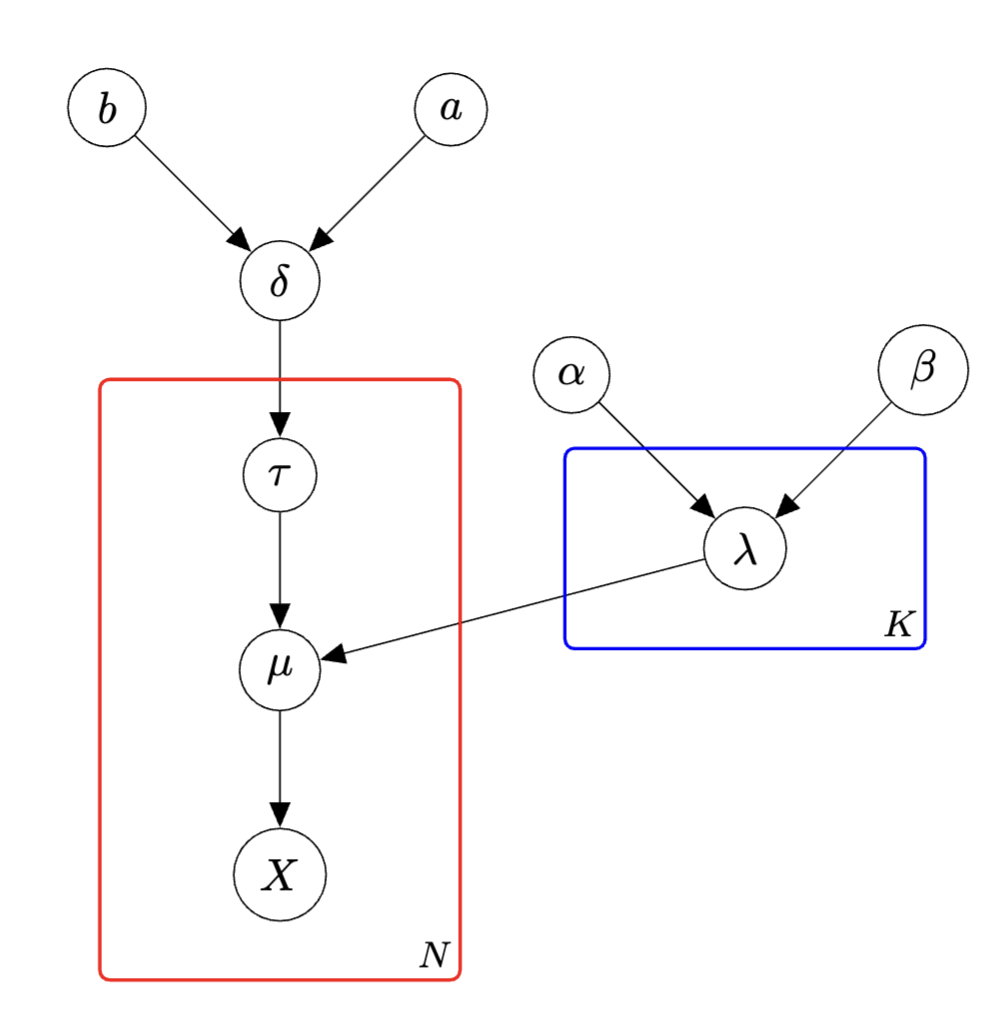}
    \caption{Graphical model of the partial membership formulation}
\label{fig1}
\end{figure}

In this work, we focus on the case when the form of the distribution for each cluster $P_{k}(\bm{x}_{i}\mid\bm{\theta}_{i})$ are independent Poisson distributions:
\begin{equation*}
    P_{k}(\bm{x}_{i}\mid\bm{\theta}_{k}) = \prod_{j=1}^{J}P_{k}(\bm{x}_{ij}\mid\lambda_{kj})=\prod_{j=1}^{J}\frac{\lambda_{kj}^{\bm{x}_{ij}}e^{-\lambda_{kj}}}{\bm{x}_{ij}!}, 
\end{equation*}
where $\bm{\theta}_k = (\lambda_{k1}, \lambda_{k2}, \ldots, \lambda_{kJ})$.
Thus, the partial membership model assumes that each data point is drawn from
\begin{equation*}
    x_{ij}\mid \bm{\tau}_i\sim \mbox{Poisson}\left[\exp\left(\sum_{k=1}^{K}\tau_{ik}\log\lambda_{kj}\right)\right].
\end{equation*}
Furthermore, we assume that the partial membership weights are drawn from a Dirichlet distribution:
\begin{equation*}
    \bm{\tau}_{i}\sim \mbox{Dirichlet}(\bm{\delta}).
\end{equation*}
We consider a model with $K$ clusters and specify prior distributions for the model parameters. In particular, we let $\bm{\delta}$ be a $K$-dimensional vector of positive hyperparameters $\bm{\delta}\sim \mbox{Unif}(a,b)$; the choice of the Uniform distribution for $\delta$ and the values of $(a,b)$ will be explained in Section~\ref{sub:identif}.
We use a gamma conjugate prior distribution for the Poisson parameters
\begin{equation*}
    P(\lambda_{kj})\propto \lambda_{kj}^{\alpha-1}e^{-\beta\lambda_{kj}},
\end{equation*}
where $\alpha$ and $\beta$ are hyperparameters of the prior. We found that the use of priors with different (sensible) choices of hyperparameters were found to have little effect on the clustering obtained in Section~\ref{sec: applBall} and Appendix~\ref{sec: applBike}. A graphical model representation of the partial membership model is given in Figure~\ref{fig1}.

In Figure~\ref{fig1}, the red plate indicates nodes repeated for each observation $i = 1, \dots, N$. Within it, $\bm{\tau}_i$ denotes the membership vector for observation $i$, drawn from a Dirichlet distribution with hyperparameter $\bm{\delta}$. The deterministic node $\bm{\mu}_i$ combines $\bm{\tau}_i$ and the cluster-specific Poisson parameters $\bm{\lambda}$ (enclosed in the blue plate), thereby defining the rates that generate the observed counts $\bm{X}_i$. Each element $\lambda_{k,j}$ in $\bm{\lambda}$ has a $\mbox{Gamma}(\alpha,\beta)$ prior, and arrows show how these parameters interact to govern the Poisson likelihood of the data. The nodes $(a,b)$ represent hyperparameters defining a $\mbox{Uniform}(0,10)$ prior for $\bm{\delta}$, while $(\alpha,\beta)$ serve as hyperparameters for the Gamma priors on $\bm{\lambda}$. Hence, all observations share a global prior structure but each observation’s partial membership vector $\bm{\tau}_i$ ultimately shapes $\mu_{i,j}$ and thus the distribution of $\bm{X}_i$.

While the Poisson independence assumption for the form of the distribution for
each cluster may appear restrictive, it keeps the model tractable and interpretable. In particular, because partial membership already introduces a continuous mixing over clusters, adding a correlated structure for the Poisson variables can complicate inference and obscure each cluster’s “mean profile.” As a result, we focus on independent Poisson likelihoods, acknowledging that more flexible covariance structures could be explored in future work.

\subsection{Mixed Membership Model}

A method for fitting mixed membership (MM) models to count data is outlined in \cite{white2016exponential}. In broad terms, an MM model stipulates that each observation (here, each player) has a membership vector $\bm{\tau}_i$ drawn from a Dirichlet distribution. However, crucially, \emph{each attribute} of observation $i$ (for instance, each type of pass or shot count) is then drawn from one of the $K$ latent clusters based on $\bm{\tau}_i$. 

Formally, once $\bm{\tau}_i$ is obtained, a latent cluster label $\bm{Z}_{ij}$ is sampled for each attribute $j$, typically via a multinomial or categorical distribution with parameters $\bm{\tau}_i$. Next, conditional on $\bm{Z}_{ij}$, the attribute $x_{ij}$ is drawn from the corresponding cluster-specific distribution $P(\cdot\mid\lambda_{Z_{ij}, j})$. This means that two attributes of the same player $i$ can come from two different clusters, reflecting a “mixed” or pattern at the attribute level.

In other words, MM models operate under the assumption that each data feature (such as pixels in an image analysis scenario) of a given item (e.g., an image) is conditionally independent, given the membership vector $\bm{\tau}_i$.

 The likelihood for an observation is given by:
\begin{equation*}
 x_{ij}\mid \bm{\tau}_i\sim\sum_{k}\tau_{ik}P({x}_{ij}\mid\lambda_{kj}) \quad \mbox{for} \quad j= 1,2, \ldots, J.
\end{equation*}
This formulation relies on the assumption of “factorization over attributes,” meaning that the likelihood of an observation is expressed as the product of likelihoods for its individual attributes, given $\bm{\tau}_i$:
\begin{equation*}
 P(x_{ij}\mid \bm{\tau}_i)=\prod_{j=1}^{J}\sum_{k=1}^{K}\tau_{ik}P({x}_{ij}\mid\lambda_{kj}).
\end{equation*}
While this simplifies the modeling process and is effective in capturing attribute-level patterns, it assumes conditional independence between attributes within an observation. 
This assumption holds if the attributes within a single observation are conditionally independent given $\bm{\tau}_i$.
For example, in applications like image analysis, where pixels can be viewed as exchangeable sub-objects, this assumption is reasonable.
In contrast, Partial Membership (PM) models relax the factorization assumption, allowing for joint modeling of attributes. This enables the PM model to account for attribute dependencies that may not align strictly with cluster-specific distributions.

The generative process for $\bm{X}$ in the partial membership model, compared to those in mixed membership and mixture models in Table~\ref{ta:comparison}.

\begin{table}[tbph]
\caption{Comparison of the data generation process for the mixture model, mixed membership model and partial membership models for count data.} \label{ta:comparison}
\begin{tabular}{lll}\hline
Mixture & Mixed Membership & Partial Membership \\\hline

 for($i$ in $1:N$) & for($i$ in $1:N$) & for($i$ in $1:N$)\\
    ~~$\bm{\tau}_i\sim \mbox{multinomial}(1, \bm{\pi})$ & ~~$\bm{\tau}_i \sim \mbox{Dirichlet}(\bm{\delta})$ & ~~$\bm{\tau}_i \sim \mbox{Dirichlet}(\bm{\delta})$\\
    ~~for($j$ in $1:J$) & 
    ~~for($j$ in $1:J$) & 
    ~~for($j$ in $1:J$) \\ 
  ~~~~  & ~~~~$\bm{Z}_{ij}\sim \mbox{multinomial}(1, \bm{\tau}_i)$ &  ~~~~$\mu_{ij}= \exp(\sum_{k=1}^{K}\tau_{ik}\log\lambda_{kj})$\\
    ~~~~$\bm{X}_{ij}\sim \mbox{Poisson}(\lambda_{\bm{\tau}_{i},j})$ & ~~~~$\bm{X}_{ij}\sim \mbox{Poisson}(\lambda_{Z_{ij},j})$ & ~~~~$\bm{X}_{ij}\sim \mbox{Poisson}(\mu_{ij})$\\\hline
\end{tabular}
\end{table}

The key difference in the MM approach is its per-attribute labeling via $Z_{ij}$: a single observation can thus exhibit multiple latent roles across its various attributes, rather than having one role (as in a standard mixture) or a single partial membership vector that weights the entire attribute vector (as in PM).

\subsection{Inference}

The PM posterior distribution takes the form:
\begin{eqnarray*}
    P(\bm{\tau}, \bm{\lambda}, \bm{\delta} \mid\bm{x}, \alpha, \beta, a, b) &\propto&  \prod_{i=1}^{n}\left[P(\bm{\tau}_i\mid\bm{\delta}) \prod_{k=1}^{K}\prod_{j=1}^{J}P_{k}(\bm{x}_{ij}\mid\lambda_{kj})^{\tau_{ik}}\right]\\
    & & \times P(\bm{\lambda}\mid\alpha,\beta) P(\bm{\delta}\mid a,b)
\end{eqnarray*}

Fitting the PM consists of inferring all unknown variables given $\bm{X}$, for which we employ a Bayesian approach using Monte Carlo Markov chain (MCMC). A notable advantage of PM over Mixed Membership (MM) models is that MM models require a discrete latent variable for each sub-object to indicate the mixture component from which it was drawn. This substantial number of discrete latent variables can make MCMC sampling in MM models significantly more challenging than in PM models.

The model was fitted using NIMBLE, a versatile framework for developing statistical algorithms for general model structures within R \citep{de2017programming, nimble-software:2024}. 
NIMBLE facilitates the definition of models as directed acyclic graphs (DAGs), which simplifies the deployment of a range of algorithms. It accommodates several MCMC techniques and provides a comprehensive array of resources for model evaluation, selection, and prediction error assessment. These features led us to prefer NIMBLE over other Bayesian MCMC software like Stan and JAGS.

\subsection{Identifiability}
\label{sub:identif}

Parameters in a model are not identified if the same likelihood function is obtained for more than one choice of the model parameters \citep{teicher1963identifiability, Gelman2013Bayesian}.
The issue of identifiability in finite mixture distributions demands careful consideration, particularly in a Bayesian context. \citet{fruhwirth2006finite} describes three sources of non-identifiability: (i) invariance to relabeling of components, (ii) non-identifiability stemming from potential overfitting, and (iii) a generic non-identifiability in certain classes of mixture distributions. Any of these can lead to poor convergence of MCMC methods.

In partial membership models, multiple sets of parameter values can again produce the same likelihood, creating further potential for non-identifiability. To address this, we leverage \emph{archetypal units}, observations with near-complete membership in a single cluster. By serving as “extreme” examples, these units effectively anchor the corners of the membership simplex, thus stabilizing cluster definitions and enhancing interpretability. 
Although our terminology refers to “archetypal units,” we do not perform a separate archetypal analysis as in \cite{cutler1994archetypal}, nor do we create “prototype” clusters at the model level (as in \cite{hou2022chimeral}). 
Rather, we identify archetypal units post hoc by selecting observations whose maximum membership weight is close to one.
Alongside identifying archetypal units, we must also handle invariance to relabeling (label switching). Among the various solutions \citep[e.g.,][]{celeux2000computational,KMurphy2020}, we address label switching post hoc via a probabilistic relabeling algorithm \citep{Sperrin2010}, implemented in the R package \textsf{label.switching}. This approach permutes cluster labels in each MCMC draw so that the final posterior summaries remain label-consistent across iterations.

In our application, archetypal units serve a dual function. First, they enhance interpretability by identifying archetypes that mark the boundaries of each cluster, ensuring that the resulting partitions remain meaningful in the context of football performance. Second, they help mitigate the non-identifiability that is common in PM models by introducing tangible constraints on the cluster definitions. Thus, identifying these archetypal units makes the PM model more practically useful, particularly in domains where clear-cut examples are important for understanding cluster structure. By anchoring the model with these near-pure members (e.g., particular football players), the model can still capture mixed attributes for other players while avoiding overly diffuse or indistinct clusters.

The selection of a uniform distribution with parameters $(0, 10)$ as the hyperprior for the $\bm{\delta}$ parameter in the Dirichlet distribution governing the mixture weights, i.e., $\tau_{i} \sim \mathrm{Dirichlet}(\bm{\delta})$, proved instrumental in enabling the identification of archetypal players while preserving interpretability in the application in Section~\ref{sec: applBall}. For completeness, we also experimented with a sparser mixture-weight matrix by using a smaller scalar parameter in the Dirichlet \citep{wang2009dirich}, but found that this increased sparsity led to a higher optimal number of clusters based on our information criterion. Consequently, the uniform distribution emerged as our preferred option, effectively promoting the detection of archetypal units while keeping the cluster count manageable, thus preserving interpretability.
Striking a middle ground, $\mbox{Uniform}(0,10)$ allowed some observations to achieve nearly complete membership in one cluster, effectively anchoring the membership simplex and thereby improving interpretability.

\subsection{Model Selection}
\label{subsec_modelselection}

According to \cite{watanabe2010asymptotic}, a statistical model is said to be \emph{regular} if the map taking parameters to probability distributions is one-to-one and if its Fisher information matrix is positive definite. If a model is not regular, then it is said to be \emph{singular}. If a statistical model contains a hierarchical structure or latent variables, then the model is generally singular.
In singular statistical models, the maximum likelihood estimator does not satisfy asymptotic normality.
Consequently, standard model selection criteria like the Akaike Information Criterion (AIC) and the Bayes Information Criterion (BIC) are not always appropriate for model selection.

WAIC (Widely Applicable Information Criterion) \citep{watanabe2010asymptotic} can be used for estimating the predictive loss of Bayesian models, using a sample from the full-data posterior, and it is applicable to non-regular models such as the PM.
In the present paper, WAIC is calculated from Equations 5, 12, and 13 in \cite{gelman2014understanding}, and it is the log pointwise predictive density minus a correction for effective number of parameters to adjust for overfitting.
According to \citet{millar2018conditional}, the \emph{marginalized} WAIC might be more accurate for choosing the correct model. 
Indeed, in our simulation (detailed below), we compare the WAIC conditional on all of the parameters (WAICc) with the WAIC marginalized for $\bm{\tau}$ (WAICm) and find that WAICm more reliably detects the true $K$ in a partial membership setting. 

In our real-data application, we report the $\text{WAIC}_m$. In mixture model application it has been marginalized over the mixing proportions $\bm{\pi}$.

We fit the model in a simulated data scenario to verify the number of times the correct model is selected by the WAIC conditional on all of the parameters ($\text{WAIC}_c$) and marginalized over $\bm{\tau}$ ($\text{WAIC}_m$).
We generated 100 random membership vectors from a Dirichlet($\bm{\delta}$) distribution with shape parameter $\bm{\delta} = (0.5, 0.5, 0.5, 0.5)$. 
Using these membership scores, we simulated 100 partial membership models with $N=200$, $J=25$, and $K=4$ to match the football players application scenario. 
Each $\lambda_{kj}$ was drawn from an $\mbox{Exponential}(1/5)$ distribution, yielding an average rate of 0.2. Additionally, each $\lambda_{kj}$ in the partial membership model was assigned a $\mathrm{Gamma}(1,1)$ prior, as this relatively uninformative choice had little effect on the estimated clusters.
We ran the MCMC algorithm for 20000 iterations, keeping every 50th draw, and discarded the first 5000 iterations as burn-in.
MCMC was run for $K=1,\dots,8$, and convergence was assessed by examining trace plots. 

The simulation results are illustrated in Table~\ref{Tab:simres}, confirming that the marginalized WAIC ($\text{WAIC}_m$) is more accurate for choosing the correct model, with 99\% of runs favoring the true $K=4$.

\begin{table}[tbph]
\centering
\caption{Number of times each $K$ yields the minimum WAIC conditional on parameters ($\text{WAIC}_c$) and marginalized over $\bm{\tau}$ ($\text{WAIC}_m$). $N=200$, $J=25$, Number of runs=100, true $K=4$. 
(Values in bold indicate the best-performing $K$.)}
\label{Tab:simres}
\begin{tabular}{rcc}
  \hline
 & $\text{WAIC}_c$ & $\text{WAIC}_m$ \\ 
  \hline
  $K=1$ & 0 & -\\
  $K=2$ & 0 & 0\\ 
  $K=3$ & 0 & 0\\ 
  $K=4$ & \textbf{79} & \textbf{99}\\ 
  $K=5$ & 16 & 1\\ 
  $K=6$ & 3 & 0\\
  $K=7$ & 2 & 0\\
  $K=8$ & 0 & 0\\
   \hline
\end{tabular}
\end{table}

Finally, we note that WAIC values from partial membership, mixed membership, and finite mixture models may appear on different numeric scales because each model specifies a distinct likelihood factorization. In Table~\ref{Tab:WAIC}, for example, the ($\text{WAIC}_m$) values for PM can be somewhat larger than for MM or mixture. However, the primary goal is to compare models within each class for a given dataset and to note which model achieves the lowest WAIC overall.

\section{Serie A Football Players}
\label{sec: applBall}

We analyze the performance statistics of 200 Serie A football players who played more than 1720 minutes during the 2022/2023 football season\footnote{The data are freely available at \url{https://fbref.com/en/comps/11/2022-2023/stats/2022-2023-Serie-A-Stats}.}.
The analysis considers a set of 22 count variables for each player across the games that they played, selected to encompass the essential skills associated with each player's role on the field. 

This application allows us to verify the reliability of the model's results by comparing them with each player's designated playing position. The player's partial membership  allows us to estimate the positions on the field where the players tend to play, in addition to their (primary) designated position, based on their playing style.

In the field of clustering football players' positions, a study with objectives similar to ours was conducted by \cite{seth2016probabilistic}, employing archetypal analysis. A key distinction between our study and that study lies in the methodology adopted, as well as the nature of the data used. \cite{seth2016probabilistic} based their analysis on skill ratings from the PES Stats Database (\url{https://www.pesmaster.com}), a community-driven platform that compiles statistics and skill ratings for soccer players. This database was originally developed for the video game “Pro Evolution Soccer” by Konami. In contrast, our study utilizes actual statistical data recorded during games, providing a different perspective and basis for analysis.

The performance variables included in the analysis and modeling are given as follows.

\begin{multicols}{2}
\centering
\begin{itemize}
    \item \textbf{Gls} -- Number of goals.
    \item \textbf{Ast} -- Number of assists. 
    \item \textbf{PrgC} -- Progressive carries: carries that move the ball towards the opponent's goal line at least 10 yards from its furthest point in the last six passes, or any carry into the penalty area. Excludes carries which end in the defending 50\% of the pitch.
    \item \textbf{Sh} -- Shots Total: does not include penalty kicks.
    \item \textbf{KP} -- Key Passes: Passes that directly lead to a shot (assisted shots).
    \item \textbf{CrsPA} -- Crosses into Penalty Area: completed crosses into the 18-yard box, not including set pieces.
    \item \textbf{SCA} -- Shot-Creating Actions: the two offensive actions directly leading to a shot, such as passes, take-ons and drawing fouls.
    \item \textbf{TO} -- SCA (TO): successful take-ons that lead to a shot attempt.
    \item \textbf{ShToSh} -- SCA (Sh): shots that lead to another shot attempt.
    \item \textbf{Def} -- SCA (Def): defensive actions that lead to a shot attempt.
    \item \textbf{GCA} -- Goal-Creating Actions: the two offensive actions directly leading to a goal, such as passes, take-ons and drawing fouls. Note: a single player can receive credit for multiple actions and the shot-taker can also receive credit.
    \item \textbf{Tkl} -- Tackles: Number of players tackled
    \item \textbf{Blocks} -- Number of times blocking the ball by standing in its path
    \item \textbf{Int} -- Interceptions
    \item \textbf{Clr} -- Clearances
\end{itemize}
\end{multicols}

Partial membership (PM), mixed membership (MM) and mixture models are applied to the dataset, with the $\text{WAIC}_m$ suggesting that the $K=4$, $K=5$ and $K=6$ profile models are optimal, as illustrated in Table \ref{Tab:WAIC}.
All three models are estimated under a Bayesian framework using the same general MCMC procedure described in Section~\ref{subsec_modelselection}. Specifically, we implement each model in Nimble, specify priors for all parameters, run a burn-in and thinning period, and then derive posterior summaries for inference. Although we summarize results (such as membership vectors or cluster parameters) via posterior means and highlight the most likely cluster assignments, these point estimates remain rooted in the Bayesian posterior. This ensures that all uncertainty quantification and model selection (based on $\text{WAIC}_m$) are conducted consistently across the three model types, while any “frequentist-looking” statistics (like means or modes) are simply convenient summaries of the posterior distribution.

\begin{table}[H]
\centering
\caption{$\text{WAIC}_m$ values (rounded to the nearest integer) for the partial membership (PM), mixed membership (MM), and mixture models, with $K$ ranging between 2 and 8. The optimal model for each model type is in bold.}
\label{Tab:WAIC}
\begin{tabular}{rcccc}
  \hline
   & $K=2$ & $K=3$ & $K=4$ & $K=5$ \\
   \hline
   PM & 46137 & 36360 & \textbf{33415} & 37796 \\
   MM & 26823  & 22702 & 21330 & \textbf{20860} \\
   Mixture & 30816 & 24648 & 23090 & 21948 \\
   \hline
   & $K=6$ & $K=7$ & $K=8$ & \\
   \hline
   PM & 34398 & 36138 & 37288 & \\
   MM & 20919 & 21422 & 21518 & \\
   Mixture & \textbf{21475} & 22015 & 22486 & \\
   \hline
\end{tabular}
\end{table}

\subsection{Partial Membership Model Results}

\begin{figure}[H]
    \centering
    \includegraphics[width=1\textwidth]{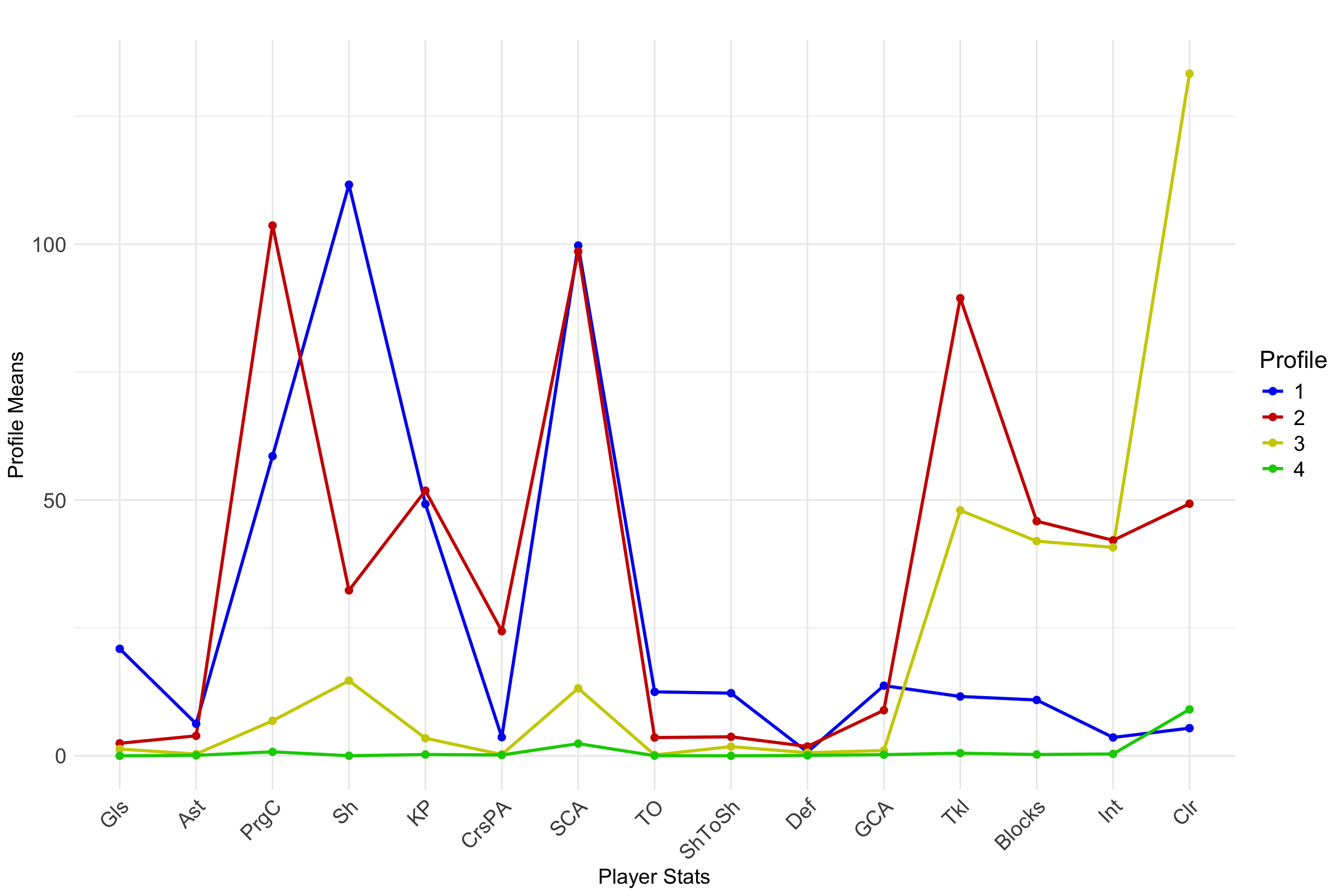}
    \caption{PM model. Expected profiles means, conditional on profile membership, with 4 profiles.}
\label{lambdahatball}
\end{figure}

Table~\ref{Tab:lambdahatball} and Figure~\ref{lambdahatball} display the posterior mean of each profile’s Poisson parameters for the partial membership (PM) model with $K=4$. The PM model allows each player to lie anywhere in the simplex of these four profiles, thus naturally capturing hybrid or multi-role behaviors.

\begin{table}[H]
\centering
\caption{PM model expected profiles means (Poisson rate parameters).}
\label{Tab:lambdahatball}
\begin{tabular}{rcccccccc}
  \hline
 & Gls & Ast & PrgC & Sh & KP & CrsPA & SCA & TO\\
  \hline
1 & 20.91 & 6.26 & 58.57 & 111.62 & 49.21 & 3.64 & 99.73 & 12.50 \\
2 & 2.44 & 3.90 & 103.65 & 32.33 & 51.82 & 24.35 & 98.56 & 3.56 \\
3 & 1.32 & 0.34 & 6.84 & 14.68 & 3.43 & 0.25 & 13.19 & 0.16 \\
4 & 0.00 & 0.09 & 0.77 & 0.01 & 0.25 & 0.15 & 2.38 & 0.02 \\
\hline
 & ShToSh & Def & GCA & Tkl & Blocks & Int & Clr \\
\hline
1 & 12.24 & 0.72 & 13.71 & 11.59 & 10.91 & 3.58 & 5.40 \\
2 & 3.71 & 1.82 & 8.90 & 89.44 & 45.86 & 42.11 & 49.27 \\
3 & 1.79 & 0.60 & 1.03 & 47.97 & 41.96 & 40.73 & 133.32 \\
4 & 0.00 & 0.07 & 0.22 & 0.51 & 0.25 & 0.36 & 9.07 \\
\hline
\end{tabular}
\end{table}

The four profiles can be interpreted as follows:

\begin{itemize}
\item \textbf{Profile 1}: 
Exhibits exceptionally high values in \emph{Shots (Sh)}, \emph{Goals (Gls)}, \emph{Assists (Ast)}, \emph{Key Passes (KP)}, and \emph{Shot-Creating Actions (SCA)}. The notable presence of \emph{Take-Ons (TO)} and \emph{Shots that lead to another shot attempt (ShToSh)} further underscores an aggressive, forward-oriented playing style. This profile aligns with \textbf{strikers} or highly offensive forwards who focus on scoring and creating chances.

\item \textbf{Profile 2}: 
Remarkably high in \emph{Progressive Carries (PrgC)}, \emph{Key Passes (KP)}, \emph{Crosses into the Penalty Area (CrsPA)}, and \emph{Shot-Creating Actions (SCA)}, yet also excels defensively with large counts of \emph{Tackles (Tkl)}, \emph{Blocks}, and \emph{Interceptions (Int)}. The presence of moderate \emph{Goals (Gls)} and \emph{Assists (Ast)} suggests a well-rounded role, typical of \textbf{full-backs and dynamic midfielders} who contribute in both offensive and defensive phases. 

\item \textbf{Profile 3}:
Dominated by very high \emph{Clearances (Clr)}, with substantial \emph{Tackles}, \emph{Blocks}, and \emph{Interceptions} as well. Offensively, it registers relatively low values in \emph{Goals} and \emph{Shots}, suggesting that these players rarely assume forward duties. These characteristics match classical \textbf{defenders}, especially center-backs, who prioritize marking, blocking, and clearing the ball.

\item \textbf{Profile 4}: 
Displays consistently low counts for nearly all offensive and defensive variables, aside from a modest number of \emph{Clearances (Clr)}. This profile is best interpreted as \textbf{goalkeepers}, who typically do not generate large numbers of shots, passes, or tackles. Clearances may occur when the keeper acts as a sweeper or clears loose balls, but otherwise these players’ statistics remain low across the board.
\end{itemize}

\begin{figure}[H]
    \centering
    \includegraphics[width=1\textwidth]{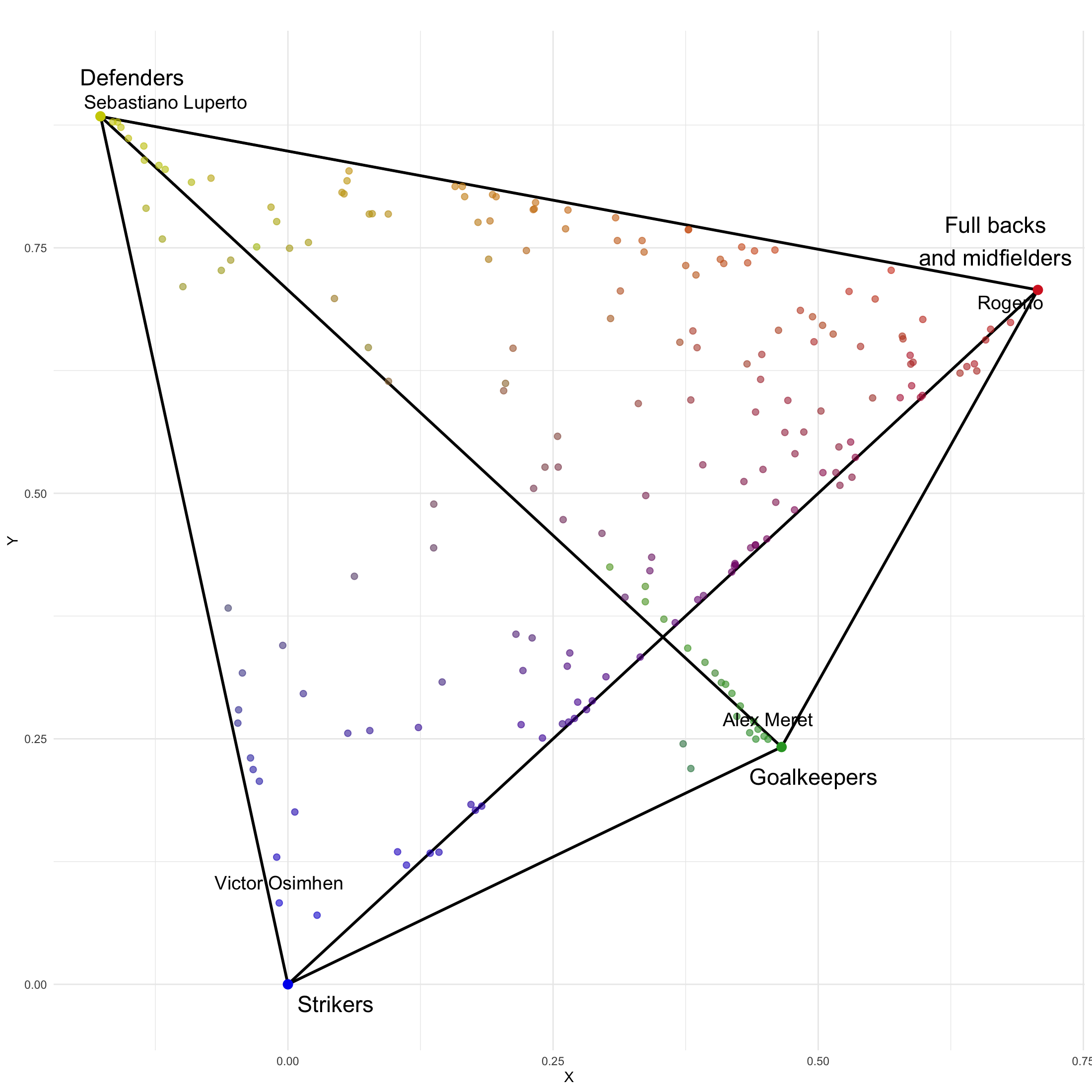}
    \caption{Tetrahedron Representation of player profiles memberships.}
\label{fig:tetrahedron}
\end{figure}

\noindent
The estimated memberships $\bm{\tau}_i$ are represented in the tetrahedron in Figure~\ref{fig:tetrahedron}. This plot provides a view of how each player’s profile splits across the four clusters: players near a corner (e.g., Victor Osimhen, Sebastiano Luperto, Rogério and Alex Meret) show almost “pure” membership in one cluster, whereas those scattered within the interior exhibit genuine multi-role behaviors.

\begin{figure}[H]
    \centering
    \includegraphics[width=1\textwidth]{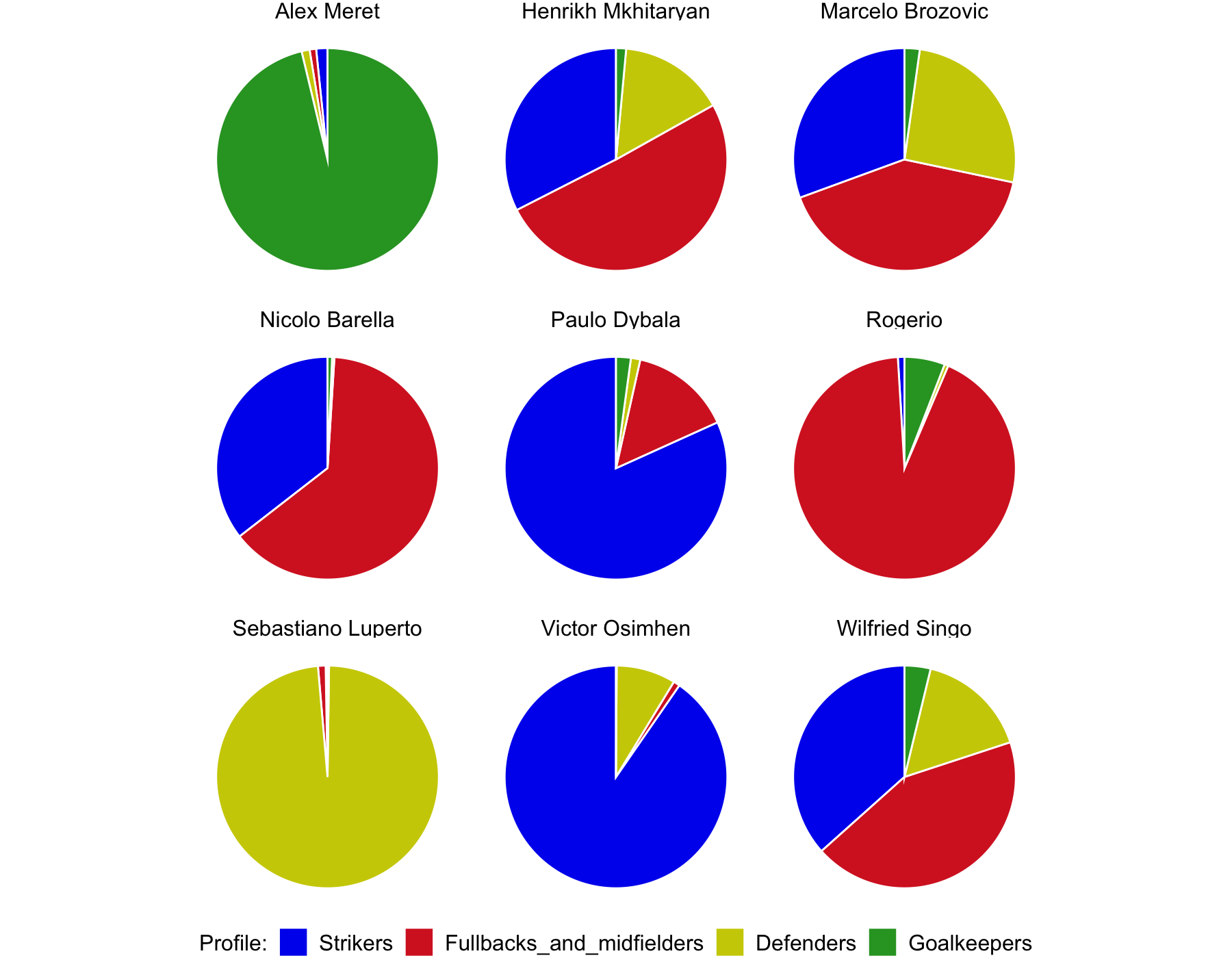}
    \caption{Partial membership model. Selected players’ membership vectors, with pie charts illustrating their fractional membership in each of the 4 profiles.}
\label{fig:taupiemainplayers}
\end{figure}

\noindent
Figure~\ref{fig:taupiemainplayers} illustrates the memberships of the archetypal players and other interesting players who split their membership. Notably:
\begin{itemize}
\item \textbf{Victor Osimhen} (Napoli) is heavily associated with Profile 1 with membership $0.904$, confirming his striker role.
\item \textbf{Rogério Oliveira da Silva} (Sassuolo) has strong membership in Profile 2, namely $0.926$, aligning with his full-back/midfielder responsibilities.
\item \textbf{Sebastiano Luperto} (Empoli) fits almost exclusively into Profile 3 with $0.984$ membership, exemplifying a traditional defender’s style.
\item \textbf{Alex Meret} (Napoli) anchors Profile 4 with $0.962$ membership, as a classic goalkeeper.
\end{itemize}
Additional players such as Paulo Dybala and Henrikh Mkhitaryan demonstrate partial memberships reflecting \emph{hybrid} roles that draw from both offensive and midfield profiles. Nicolò Barella’s split between Profiles 1 and 2 illustrates a box-to-box midfielder with notable attacking flair, while Marcelo Brozovic’s combination of Profiles 2, 1, and 3 underscores his deep-lying playmaking and defensive screening abilities.
Lastly, Wilfried Singo, known for his physical strength and speed as a right full-back or outside midfielder, is characterized by his dynamic play, adept in both offensive drives and defensive recoveries.
As illustrated in Figure~\ref{fig:taupiemainplayers}, the partial membership model encapsulates these diverse characteristics of the players well, demonstrating its efficacy in capturing the complex nature of the roles and abilities of football players. The expected values of $\tau$, estimated as posterior means of the partial membership weights, represented for nine players in Figure~\ref{fig:taupiemainplayers}, summarize the relative contributions of each cluster to a player's overall profile. For example, a player with $\tau=(0.25,0.25,0.25,0.25)$ is interpreted as balancing characteristics between four roles, rather than belonging exclusively to one with uncertainty.
The resulting insights can be valuable for coaches, scouts, and analysts who wish to understand and leverage the multi-faceted contributions of each player on the pitch.

\subsection{Mixed Membership Model Results}

\begin{figure}[H]
    \centering
    \includegraphics[width=1\textwidth]{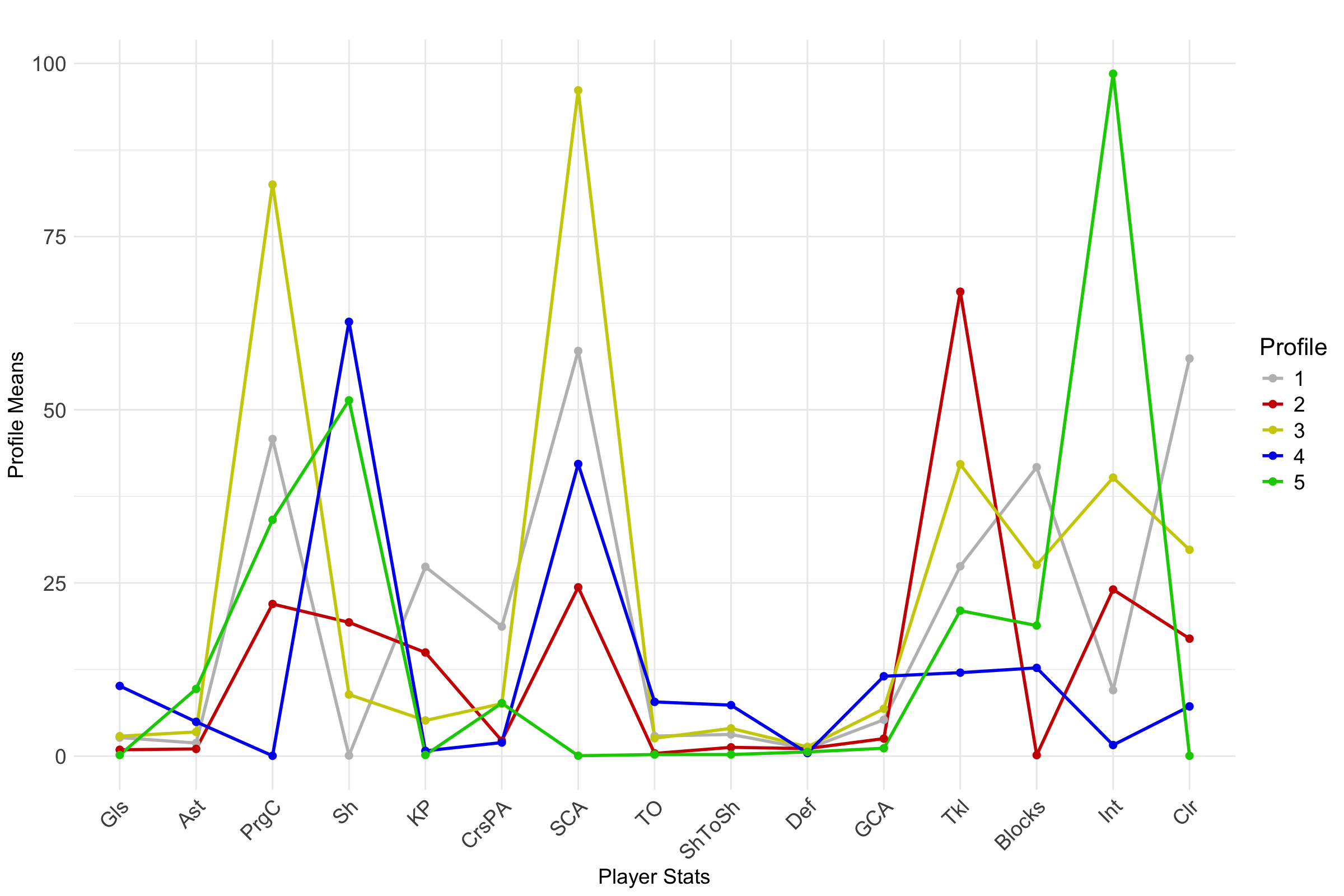}
    \caption{MM model. Expected profile means, conditional on profile membership.}
\label{lambdahatballmixed}
\end{figure}

Mixed membership (MM) models for count data are discussed in \cite{white2016exponential}, and a detailed description of their structure is given therein. We fit an MM model to the same Serie A player data under consideration. The estimated Poisson means for each of the five profiles are shown in Table~\ref{Tab:lambdahatballmixed} and visualized in Figure~\ref{lambdahatballmixed}. 

Although the MM framework allows an observation’s \emph{individual attributes} (e.g.\ passes, tackles) to be drawn from different clusters, it does not identify ``archetypal’’ players who fully belong to a single role (as partial membership models might). In practice, this means the MM model may capture hybrid behaviors, but interpreting each profile can be less straightforward than with partial membership. 

\begin{table}[H]
\centering
\caption{Mixed membership model: expected profile means (Poisson rate parameters).}
\label{Tab:lambdahatballmixed}
\begin{tabular}{rcccccccc}
  \hline
   & Gls & Ast & PrgC & Sh & KP & CrsPA & SCA & TO \\
  \hline
1 & 2.68 & 1.89 & 45.78 & 0.09 & 27.33 & 18.70 & 58.49 & 2.90 \\
2 & 0.92 & 1.04 & 21.96 & 19.31 & 14.96 & 2.22 & 24.37 & 0.39 \\
3 & 2.88 & 3.49 & 82.51 & 8.89 & 5.15 & 7.58 & 96.11 & 2.58 \\
4 & 10.13 & 4.95 & 0.05 & 62.69 & 0.76 & 1.96 & 42.17 & 7.83 \\
5 & 0.16 & 9.70 & 34.11 & 51.36 & 0.17 & 7.65 & 0.06 & 0.23 \\
\hline
 & ShToSh & Def & GCA & Tkl & Blocks & Int & Clr \\
\hline
1 & 3.13 & 1.07 & 5.25 & 27.41 & 41.71 & 9.52 & 57.39 \\
2 & 1.27 & 1.09 & 2.53 & 67.04 & 0.14 & 24.05 & 16.96 \\
3 & 4.01 & 1.34 & 6.84 & 42.14 & 27.60 & 40.21 & 29.79 \\
4 & 7.36 & 0.44 & 11.53 & 12.05 & 12.74 & 1.60 & 7.18 \\
5 & 0.23 & 0.61 & 1.14 & 21.00 & 18.85 & 98.51 & 0.05 \\
\hline
\end{tabular}
\end{table}

We now provide a tentative interpretation of each profile based on the most prominent variables:

\begin{itemize}
    \item \textbf{Profile 1.} Moderately high \emph{progressive carries} (PrgC) and notably large values for \emph{shot-creating actions} (SCA) and \emph{clearances} (Clr). The combination of moderate offensive attributes (KP, Ast) with strong defensive indicators (Blocks, Tkl, Clr) could represent wide defenders or \emph{full-backs} who frequently move the ball forward and also clear threats.

    \item \textbf{Profile 2.} More balanced attacking contributions (Sh~$= 19.31$, KP~$= 14.96$) but a standout \emph{tackle} count (Tkl~$= 67.04$) and significant \emph{interceptions} (Int~$= 24.05$). This profile suggests a \emph{defensive-minded midfielder}, someone who can take a moderate number of shots but excels at reclaiming possession and breaking up play.

    \item \textbf{Profile 3.} Very high \emph{progressive carries} (82.51) and \emph{SCA} (96.11), combined with substantial defensive metrics (Tkl~$= 42.14$, Blocks~$= 27.60$, Int~$= 40.21$). This cluster may reflect \emph{all-around midfielders} or possibly wide players who contribute heavily to both ball progression and defense, to box-to-box midfielders or very active wing-backs.

    \item \textbf{Profile 4.} Strikingly high \emph{goals} (10.13), \emph{shots} (62.69), and \emph{shot-to-shot passes} (ShToSh $= 7.36$), but minimal progressive carries (0.05). This points to a \emph{pure striker or center-forward} role, someone who focuses on finishing rather than ball-carrying.

    \item \textbf{Profile 5.} Low scoring (Gls~$=0.16$) but a large number of \emph{assists} (9.70), many \emph{shots} (51.36), and an extremely high \emph{interceptions} count (Int~$= 98.51$). This combination might characterize a \emph{deep-lying playmaker} or \emph{attacking midfielder} who also excels at intercepting passes, suggesting an unusual hybrid of offensive and defensive ability.
\end{itemize}

Although these labels offer a rough guide, the mixed membership model essentially spreads each player’s attributes across multiple latent profiles. Consequently, each of these profiles may mix features of different roles, making them more challenging to interpret than profiles from the partial membership model. Full membership values for each player are provided in the supplementary materials.

\subsection{Finite Mixture Model Results}

Mixture models for count data have been applied in contexts similar to ours \citep[e.g.,][]{white2016exponential}, often to compare with mixed membership approaches. For a foundational explanation of mixture modeling, we refer to \cite{Everitt1981} and \cite{Bouveyron2019Jun}.

\begin{figure}[H]
    \centering
    \includegraphics[width=1\textwidth]{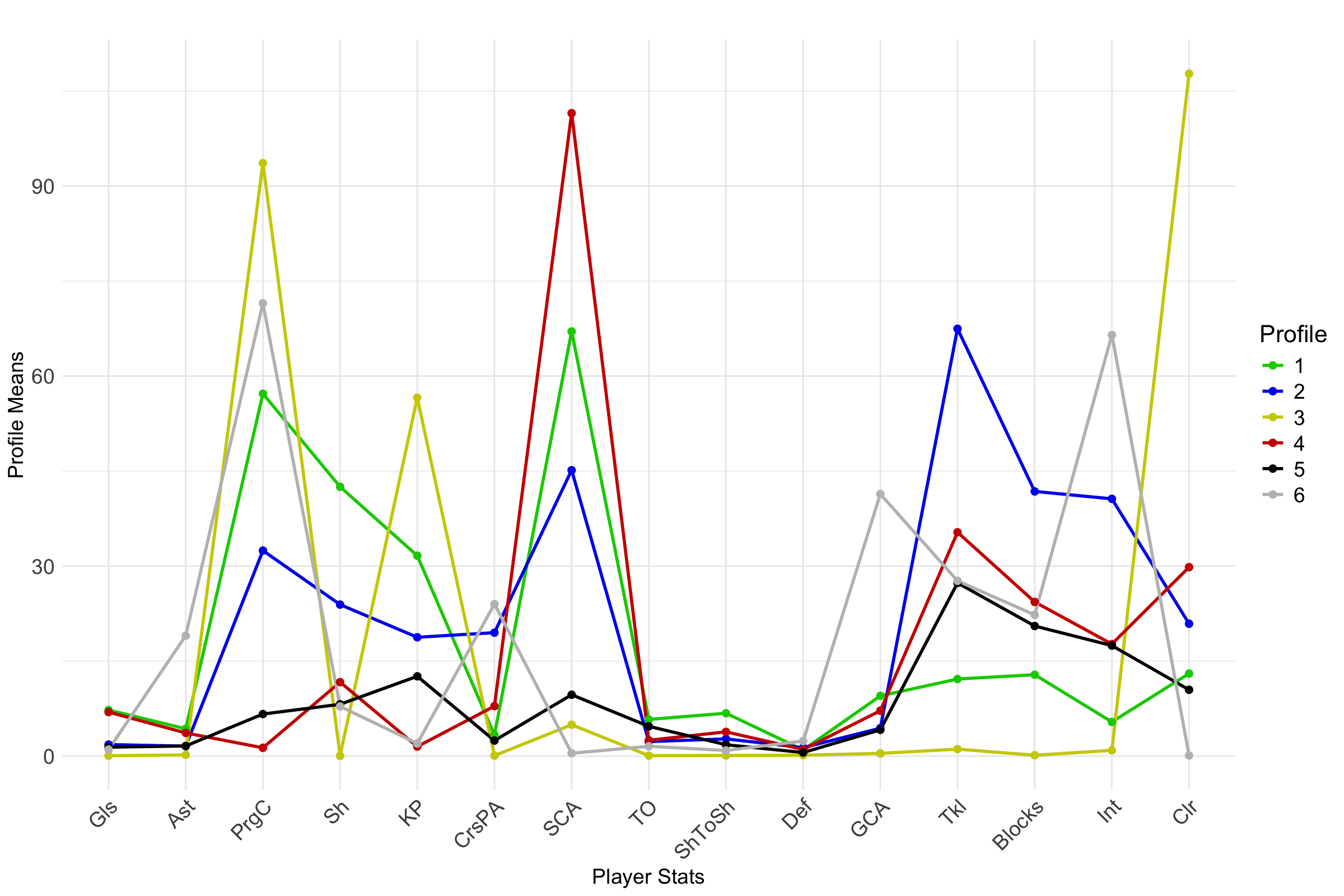}
    \caption{Finite mixture model with 6 components. Poisson means for each variable are plotted for each component.}
\label{lambdahatballmixture}
\end{figure}

We fit a multivariate Poisson mixture model to the player data. The component-specific means are shown in Table~\ref{Tab:lambdahatballmixture} and illustrated in Figure~\ref{lambdahatballmixture}. Because finite mixture models assign each player to exactly one component, they do not explicitly accommodate players who \emph{truly} balance multiple roles. Instead, they treat multi-role players as uncertain assignments to a single “best-fitting” component.

\begin{table}[H]
\centering
\caption{Finite mixture model: expected component means (Poisson rate parameters).}
\label{Tab:lambdahatballmixture}
\begin{tabular}{rcccccccc}
\hline
 & Gls & Ast & PrgC & Sh & KP & CrsPA & SCA & TO \\
  \hline
1 & 7.00 & 5.44 & 71.51 & 55.83 & 48.87 & 9.98 & 97.65 & 6.12 \\
2 & 0.98 & 1.37 & 29.07 & 16.93 & 15.53 & 3.99 & 40.32 & 0.69 \\
3 & 0.05 & 0.23 & 0.05 & 0.05 & 0.71 & 0.05 & 5.05 & 0.05 \\
4 & 1.15 & 0.55 & 11.09 & 11.76 & 4.69 & 0.75 & 15.62 & 0.23 \\
5 & 2.43 & 2.60 & 61.94 & 27.44 & 31.13 & 9.19 & 69.52 & 2.93 \\
6 & 2.48 & 35.57 & 52.53 & 24.10 & 1.84 & 56.84 & 4.17 & 5.74 \\
\hline
 & ShToSh & Def & GCA & Tkl & Blocks & Int & Clr \\
\hline
1 & 6.12 & 0.90 & 11.21 & 30.24 & 21.40 & 12.90 & 13.98 \\
2 & 1.93 & 1.03 & 3.34 & 55.80 & 35.20 & 32.76 & 60.96 \\
3 & 0.05 & 0.09 & 0.52 & 1.01 & 0.15 & 0.67 & 12.51 \\
4 & 1.29 & 0.53 & 1.36 & 37.63 & 31.98 & 29.42 & 93.07 \\
5 & 3.23 & 1.30 & 6.58 & 50.16 & 29.04 & 23.19 & 35.71 \\
6 & 0.83 & 6.43 & 15.43 & 13.97 & 5.45 & 17.09 & 0.13 \\
\hline
\end{tabular}
\end{table}

Based on the most prominent variables, we interpret the six components as follows:

\begin{itemize}
\item \textbf{Component 1}:
   Demonstrates notably high \emph{Goals} (7.00), \emph{Shots} (55.83), \emph{Key Passes} (48.87), and \emph{Shot-Creating Actions} (97.65). 
   Though there is also a moderate defensive contribution (e.g., Tkl~$= 30.24$, Int~$= 12.90$), its standout offensive metrics suggest an \emph{attacking-focused role}, such as a forward or advanced midfielder who both scores and sets up goals.

\item \textbf{Component 2}:
   Balances moderate offense (Shots~$= 16.93$, SCA~$= 40.32$) with robust defensive numbers (Tkl~$= 55.80$, Blocks~$= 35.20$, Int~$= 32.76$, Clr~$= 60.96$). 
   This dual profile points to \emph{two-way or defensive-minded midfielders} capable of contributing in attack but primarily excelling at regaining possession and thwarting opponents.

\item \textbf{Component 3}:
   Characterized by very low values across most attacking and defensive actions (e.g., Goals~$= 0.05$, Shots~$= 0.05$, Tkl~$= 1.01$, Blocks~$= 0.15$). However, Clearances~$= 12.51$ is non-negligible. 
   These sparse counts could indicate \emph{goalkeepers} or rare outliers who make few direct contributions in open play but occasionally clear the ball.

\item \textbf{Component 4}:
   Dominated by extremely high \emph{Clearances} (93.07), combined with substantial \emph{Tackles} (37.63), \emph{Blocks} (31.98), and \emph{Interceptions} (29.42). This defensive emphasis overshadows a moderate attacking presence (Shots~$= 11.76$, SCA~$= 15.62$). 
   Consequently, this profile aligns with \emph{classic center-backs or defense-first players} who make frequent clearances and recoveries but still might push forward occasionally.

\item \textbf{Component 5}:
   Displays considerable \emph{Progressive Carries} (61.94), \emph{Key Passes} (31.13), and \emph{Tackles} (50.16), along with moderate shooting (Shots~$= 27.44$). Its overall mix suggests players who energetically move the ball forward while contributing in defense. 
   This could fit \emph{full-backs or box-to-box midfielders} who link offense and defense through carries, passing, and ball recoveries.

\item \textbf{Component 6}:
   Features very high \emph{Assists} (35.57) and \emph{Crosses into the Penalty Area} (56.84), indicating a strong creative and wide-play dimension. It also has a moderate share of \emph{Shots} (24.10) and \emph{Interceptions} (17.09). 
   These attributes align with \emph{attacking wing-backs or wide midfielders} who frequently deliver crosses, set up goals, and offer some defensive coverage.
\end{itemize}

Overall, the finite mixture model reveals distinct positional tendencies but, unlike the partial membership framework, does not allow for explicit nuanced roles. Each player is assigned to one of these six clusters according to a posterior probability that captures uncertainty rather than genuine mixed roles. Consequently, for players who genuinely balance multiple responsibilities, interpretation can be less direct than in the PM model. 
A partial membership framework (PM) may offer clearer insight for players whose roles genuinely span multiple clusters.

\subsection{Model Comparison}
\label{susec: models comparison}

We compare the partial membership (PM), mixed membership (MM), and finite mixture models on interpretability, computational cost, and alignment with known football roles. All three methods reveal at least one offensive group (strikers), one defensive group (defenders), and one or more midfield/transition groups, a structure consistent with standard football positions \citep[see][for a broader discussion on cluster interpretability]{FraleyHowManyClusters, forgy1965cluster}.

A key difference from MM is that the PM model naturally identifies an \emph{archetypal} (near-pure) player for each cluster, even when we fix the same number of clusters as the other models. For instance, with $K=4$, the MM model yields the highest membership values of only 0.565, 0.686, 0.576, and 0.656 for its four clusters. These do not indicate true “corner” players who fully exemplify one role. Consequently, such clusters can be harder to label consistently, especially in the finite mixture model, where combining roles in a single cluster can obscure a player’s actual multi-role tendencies. The MM model can capture hybrid behaviors at an attribute level but lacks straightforward “archetypal” players who neatly define each cluster.  

By contrast, the PM model’s continuous membership vectors, along with near complete memberships for certain archetypal players, make it more intuitive to map each cluster to a well-known role, such as a striker or a goalkeeper. MM does allow attribute-level mixing, but this can complicate the direct labeling of entire clusters. The finite mixture model, with strictly binary (one-cluster) memberships, often ends up blending distinct roles within a single component.  

Computationally, we ran each method for \(K = 2\) through \(K = 8\) under the same conditions on a MacBook Air (Apple M3, 16GB RAM) with no other processes running. The partial membership (PM) model took about 38{,}129 seconds (\(\sim\)10.6 hours), the mixed membership (MM) model 52{,}797 seconds (\(\sim\)14.7 hours), and the finite mixture model 26{,}320 seconds (\(\sim\)7.3 hours). These times reflect both the dimensionality of each model’s latent variables and the efficiency of our MCMC implementation. 

In particular, PM simultaneously samples a continuous membership vector \(\bm{\tau}_i\) for each player, while MM requires a potentially high-dimensional set of attribute-level latent labels and can thus converge more slowly. The finite mixture model here follows a “one-cluster-per-player” scheme, which substantially reduces the number of latent indicators relative to MM’s attribute-level labels; as a result, it finishes more quickly than the MM model despite having the same overall range of \(K\). In practice, exact run times depend on convergence thresholds, initialization, and hyperparameter settings, so these values should be viewed as indicative rather than absolute.

\section{Conclusions And Future Developments}
\label{sec: conclusions}

Partial membership models offer analysts significantly greater flexibility compared to traditional model-based clustering or standard distance-based clustering methods. In our study, we tailored the model specifically for count data and applied it to the analysis of Serie A football players during the 2022/2023 season. Our goal was to estimate the various positions players tend to occupy on the field, in addition to their primary positions, by analyzing their playing styles. We based our analysis on a set of 15 count variables recorded during games, carefully selected to cover the key skills pertinent to each player’s role.
This application also allowed us to test the model’s reliability by comparing its results with each player's actual playing position. Furthermore, when compared with mixed membership and mixture models for count data, the partial membership model yielded more realistic and interpretable results. It excelled particularly in capturing archetypal players, whose distinct characteristics define specific clusters. These archetypes not only aid in clarifying each cluster’s definition but also contribute to the model's overall stability and interpretability. By serving as benchmarks, they assist in better identifying and understanding the roles of players whose classifications might otherwise be ambiguous.
While the partial membership model does require considerable computational resources, we believe its potential applications extend beyond sports analysis, offering valuable insights in fields like social sciences, genetics, natural sciences, and textual analysis. It can address some limitations inherent in mixed membership models. 

In the realm of sports analytics, our findings could significantly benefit coaches, talent scouts, team managers, and analysts. Utilizing these insights, they can make more informed decisions regarding team strategy, talent acquisition, and statistical research, thereby enhancing both performance and understanding in football.

Looking ahead, two primary areas present intriguing opportunities for enhancing our model. Firstly, addressing the issue of over-dispersion, which is a common challenge in count data, is of considerable interest. Developing methods to accurately assess and incorporate over-dispersion into the model would enable more precise and reliable predictions. Secondly, adding a temporal dimension to the model opens up another avenue for exploration. By incorporating a temporal aspect, the model could provide a dynamic view of clusters and allow for more nuanced analyses. In this direction, one promising line of research would be to extend our approach to multiple football seasons and focus specifically on outfield players, thus offering deeper longitudinal insights while potentially refining the role-based clusters.

\section*{Acknowledgements}

Thomas Brendan Murphy's research was supported by the Science Foundation Ireland Insight Research Centre (12/RC/2289\_P2) and a visiting period at Collegium de Lyon.

\bibliography{MAIN_REVIEWED}%

\newpage
\appendix

\section{Application: Washington DC Bike Data}
\label{sec: applBike}

\begin{figure}[H]
    \centering
    \includegraphics[width=1\textwidth]{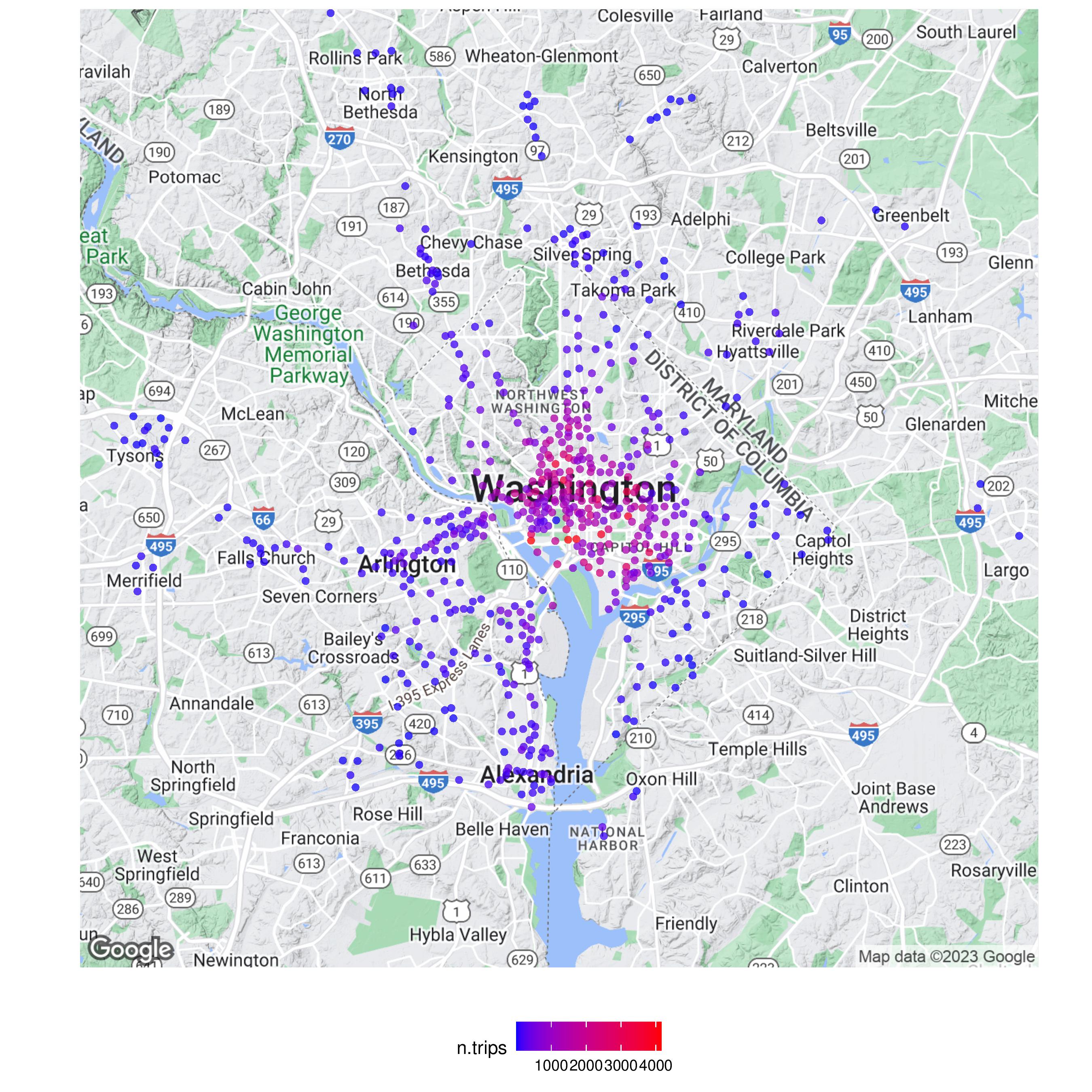}
    \caption{Number of bicycle trips per bike sharing station from 15 of June to 15 of July 2022}
\label{ntrips}
\end{figure}

We apply partial membership model on the data of the bike sharing company of Washington DC.
The data are collected daily, from June 15th to July 15th, 2022, and record each single ride: date and time of start of trip, date and time of end of trip, name, ID, longitude and latitude of starting station, name, ID, longitude and latitude of ending station\footnote{ The data are freely available at \url{https://capitalbikeshare.com/system-data}}. Figure~\ref{ntrips} shows the number of trips per station in the considered time period.

\begin{figure}[H]
    \centering
    \includegraphics[width=1\textwidth]{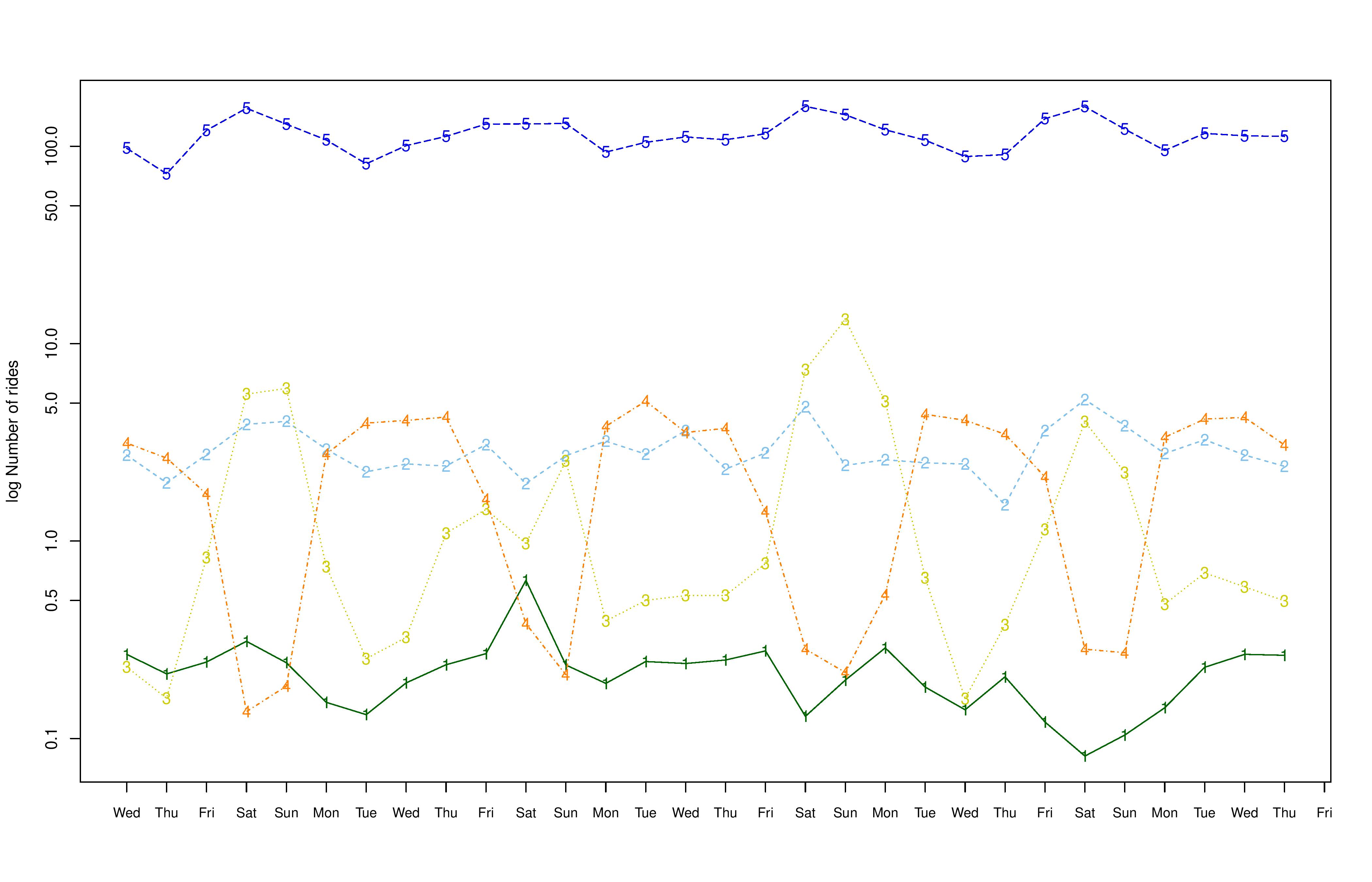}
    \caption{Log of the expected number of rides per day from June 15th to July 15th, 2022, conditional on profile membership, with 5 profiles.}
\label{lambdahat}
\end{figure}

\begin{figure}[H]
    \centering
    \includegraphics[width=1\textwidth]{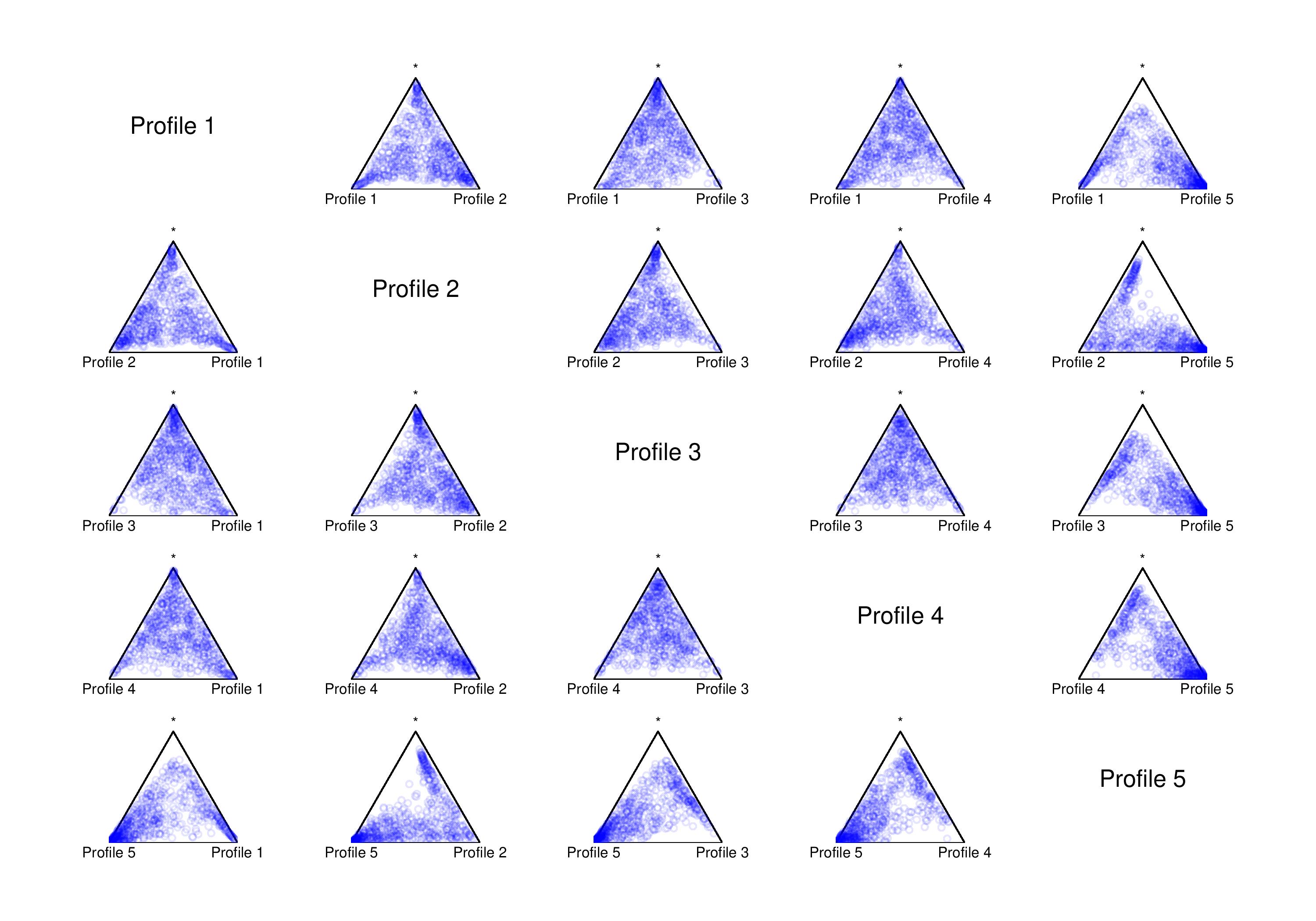}
    \caption{Marginal simplices representing bikes stations’ profile membership}
\label{tauhat}
\end{figure}

We calculated the number of times bikes are collected from each of the 660 stations and we modelled these counts using a partial membership model, with the intent to explore the interactions between the bikes stations usage, to improve the allocation of the bikes.
Partial membership model suits this type of application because the bikes move between the stations along the day, so the usage of the stations could vary and their membership could be partial. 
While one station may serve predominantly weekday commuters and another may cater mostly to weekend leisure riders, many stations show a blend of both roles (e.g., moderately busy on weekdays, highly used on weekends). By allowing “partial” membership in multiple clusters, our model better reflects these overlapping behaviors than a strictly single-cluster assignment, providing a richer picture of station usage variability.
It should be noted that we do not addressed the temporal dependency as the temporal nature of the data would require. Nevertheless, the approaches appear to identify interesting behaviour in the data, and serve to illustrate the usefulness of the method.
We run the model over a range of $K=1,\dots,6$. The model with the lowest WAIC is the one with 5 profiles (or components). For a better visualization, in Figure~\ref{lambdahat} are represented the natural log of the profiles means, while Figure~\ref{tauhat} shows the marginal simplices representing stations' profile membership.

\begin{figure}[H]
    \centering
    \includegraphics[width=1\textwidth]{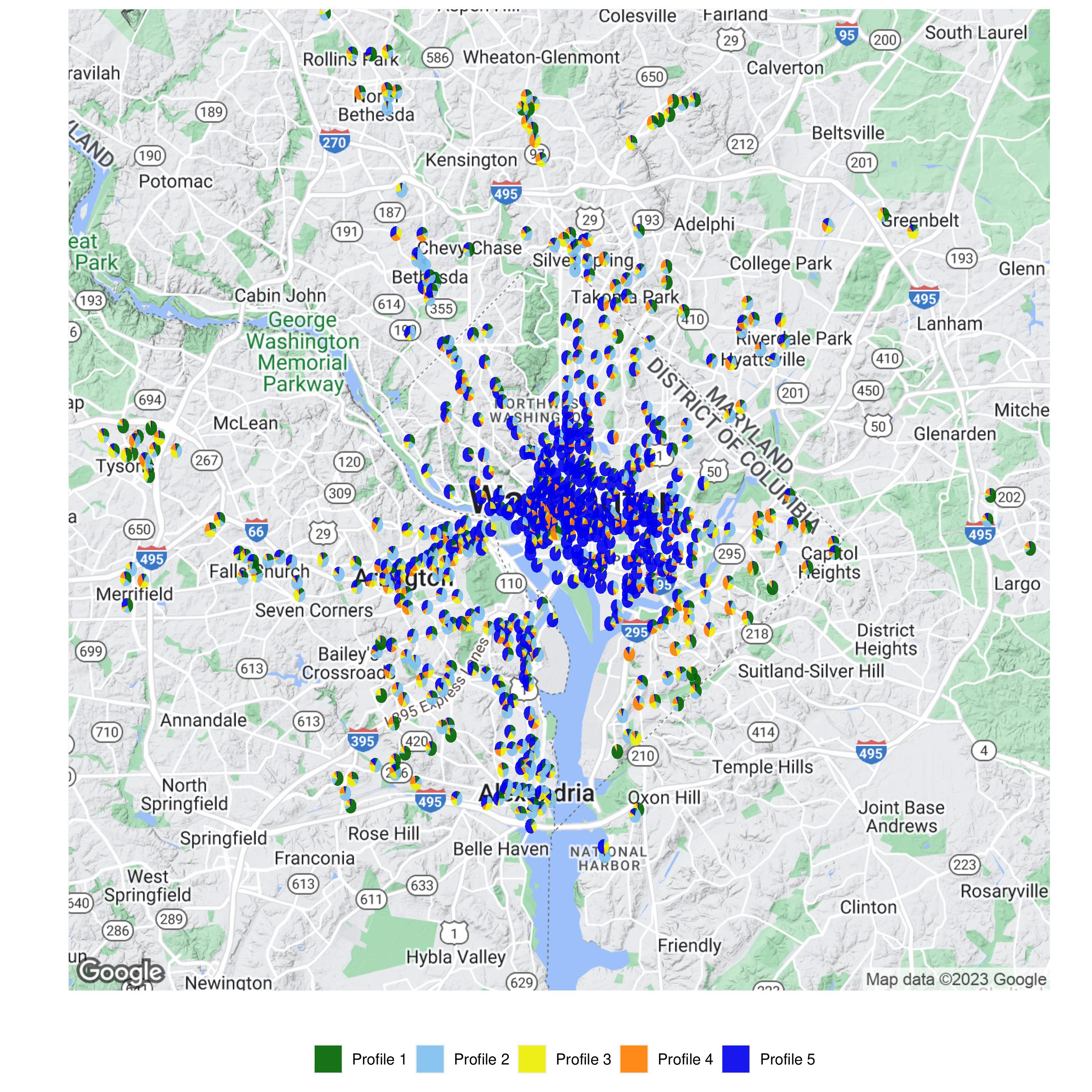}
    \caption{Bike stations' pie charts of profiles membership.}
\label{map}
\end{figure}

Figure~\ref{map} represents as a pie chart the profile membership for each of the bike stations.

It could be seen that Profile 5 groups the busiest stations, which are mainly located in the center of the city. Profile 1 the less used ones, which are mainly in the outlying areas, Profile 2 is an average usage stations cluster and looking at the map, it seems to connect the centre to the peripheral areas, Profile 3 groups the stations mostly used during the weekends, with an high peak of usage during the holiday of July 4th (Monday), which is public holiday in the USA. The stations with an high membership to this profile, are often located near the river or green ares, or also in the outlying areas. Profile 4 is the group of the stations mostly used on working days.

\end{document}